\documentclass[
reprint,
superscriptaddress,
amsmath,amssymb,
aps,
floatfix,
]{revtex4-2}
\usepackage{mathbbol}
\usepackage{amssymb}
\DeclareSymbolFontAlphabet{\amsmathbb}{AMSb}
\usepackage{graphicx}
\usepackage{dcolumn}
\usepackage{comment}
\usepackage{bm}
\usepackage{xcolor}
\usepackage{soul}
\def\be{\begin{equation}}
\def\ee{\end{equation}}
\def\la{\langle}
\def\ra{\rangle}
\newcommand{\ket}[1]{\vert#1\ra}
\newcommand{\bra}[1]{\la#1\vert}

\begin{document}

\preprint{APS/123-QED}

%\title{Optimal photonic gates for quantum-enhanced telescopes}

\title{Optimal qubit circuits for quantum-enhanced telescopes}

\author{Robert Czupryniak}
\email{rczupryn@ur.rochester.edu}
\affiliation{Department of Physics and Astronomy, University of Rochester, Rochester, NY 14627}
\affiliation{Center for Coherence and Quantum Optics, University of Rochester, Rochester, NY 14627}
\author{John Steinmetz}
\affiliation{Department of Physics and Astronomy, University of Rochester, Rochester, NY 14627}
\affiliation{Center for Coherence and Quantum Optics, University of Rochester, Rochester, NY 14627}
\author{Paul G. Kwiat}
\affiliation{Department of Physics, University of Illinois Urbana-Champaign, Urbana, IL 61801}
\affiliation{Illinois Quantum Information Science and Technology (IQUIST) Center, University of Illinois Urbana-Champaign, Urbana, IL 61801}
\author{Andrew N. Jordan}
\affiliation{Department of Physics and Astronomy, University of Rochester, Rochester, NY 14627}
\affiliation{Center for Coherence and Quantum Optics, University of Rochester, Rochester, NY 14627}
\affiliation{Institute for Quantum Studies, Chapman University, Orange, CA 92866}

\begin{abstract}
We propose two optimal phase-estimation schemes that can be used for quantum-enhanced long-baseline interferometry. By using distributed entanglement, it is possible to eliminate the loss of stellar photons during transmission over the baselines. The first protocol is a sequence of gates using nonlinear optical elements, optimized over all possible measurement schemes to saturate the Cramér-Rao bound. The second approach builds on an existing protocol, which encodes the time of arrival of the stellar photon into a quantum memory. Our modified version reduces both the number of ancilla qubits and the number of gate operations by a factor of two. 
\end{abstract}

\maketitle

\section{Introduction}

Classical long-baseline interferometry has become a widely accepted method of determining stellar distances or imaging light sources \cite{lawson2000principles,J_Monnier_review}. The central idea is to measure the coherence of the starlight incident at two or more telescopes as a function of their separation, then use the van Cittert--Zernike theorem~\cite{van_Cittert_Zernike_1, van_Cittert_Zernike} to extract information about the source. This has led to many significant advances, including the first observation of a black hole using radio telescopes~\cite{black_hole_image, wielgus2020monitoring}, exoplanet angular diameter estimation \cite{exoplanet_diameter}, and pulsar proper motion measurements \cite{deller2019microarcsecond}. However, there are fundamental limits to such classical interferometric techniques in the optical frequencies, such as quantum shot noise \cite{shot_noise} and stellar photon loss during transmission through the long baselines.

Quantum-enhanced telescopy aims to overcome these difficulties by employing concepts from quantum information theory \cite{Nielsen2000}, some of which have been implemented in experiment, including long-distance entanglement distribution \cite{1200km_entanglement, hensen2015loophole}, quantum logic gates \cite{gates1, gates2} and quantum memories \cite{memory, DLCZ_memory}. Therefore, it became attractive to design interferometric setups using these quantum resources. The development of quantum repeaters \cite{Sangouard_repeaters_review, Simon_repeater} motivated the exploration of nonlocal setups to enable reliable, long-distance distribution of entangled quantum states.
The assumption of having long-distance entanglement as a resource was explored in several spatially nonlocal schemes of quantum-enhanced telescopy \cite{Gottesman2012, Khabiboulline, khabiboulline2019quantum}.
A \textit{spatially local scheme} for a pair of telescopes does not allow bringing the light collected by the telescopes physically together or distributing entangled quantum states between the telescope locations. For weak thermal light sources such as starlight, spatially local schemes such as heterodyne detection will always provide less information about the source when compared to the nonlocal proposals \cite{Tsang2011}. Therefore, there has been increased interest in nonlocal schemes.

The first nonlocal scheme was given by Gottesman \textit{et al.}~\cite{Gottesman2012}. They suggested the pioneering proposal of overcoming the problem of transmission losses in the long baselines by establishing a quantum repeater link \cite{Sangouard_repeaters_review} between the telescopes, but this scheme requires a high rate of entanglement distribution, making it experimentally challenging. Essentially, one needs a distributed photon ready to interfere with every possible spectral-temporal mode of the starlight, which is extremely inefficient since nearly all these modes are unoccupied. Khabiboulline \textit{et al.} \cite{Khabiboulline, khabiboulline2019quantum} showed that one can significantly reduce the needed rate of entanglement generation by implementing local quantum processing with appropriate quantum memories \cite{kimble2008quantum}. In the conceptually simplest scheme, they effectively proposed a quantum nondemolition measurement that identifies which spectral-temporal mode contains a stellar photon, without determining which telescope received the photon.

In this paper, we introduce two optimal phase estimation schemes that can be applied to long-baseline interferometry. We describe the general two-telescope setup and define what makes a measurement scheme optimal. We focus on two classes of protocols: unary protocols, where for each run of the measurement setup one has access only to the quantum state provided by the stellar source within a single time bin, and nonunary protocols, where for each run of the experiment one has access to the state provided by the stellar source across multiple time bins. The nonunary protocols often operate on the assumption that at most one of these time bins is occupied, and we will also make this assumption here. The unary protocols will likely be more feasible to demonstrate experimentally in the near future (such as in the laboratory with an artificial source imitating the star or by looking at a fairly bright astronomical object) as they only require the manipulation of the stellar state incoming within a single time bin. However, in order to work with all the available stellar photons (which only sparsely occupy the arriving modes), they require an extremely high entanglement distribution rate. The nonunary protocols have the advantage of requiring a significantly lower entanglement distribution rate, but they require more complicated quantum operations in order to manipulate the incoming stellar state for multiple time bins. Since both classes of protocols come with advantages and disadvantages, we consider improvements to both of them.

For unary protocols, we show how the idea of Gottesman \textit{et al.} can be altered to improve the precision of the phase estimate by a factor of two by using nonlinear gate operations.  For nonunary protocols, we consider a modification to the Khabiboulline \textit{et al.} scheme that reduces the number of required resources and quantum operations by half. Both proposed protocols can be used to determine the time of arrival of the star photon, while keeping the which-path information ambiguous. This is important since projecting a given stellar photon to a single path results in losing all the interference and, hence, all information about the visibility, which is the parameter we want to estimate. It is also possible to merge both of our protocols to achieve a significantly lower error rate in the unary protocol, as demonstrated in Appendix~\ref{app:gate-errors}.

\begin{figure}
    \centering
    \includegraphics[width=8.6cm]{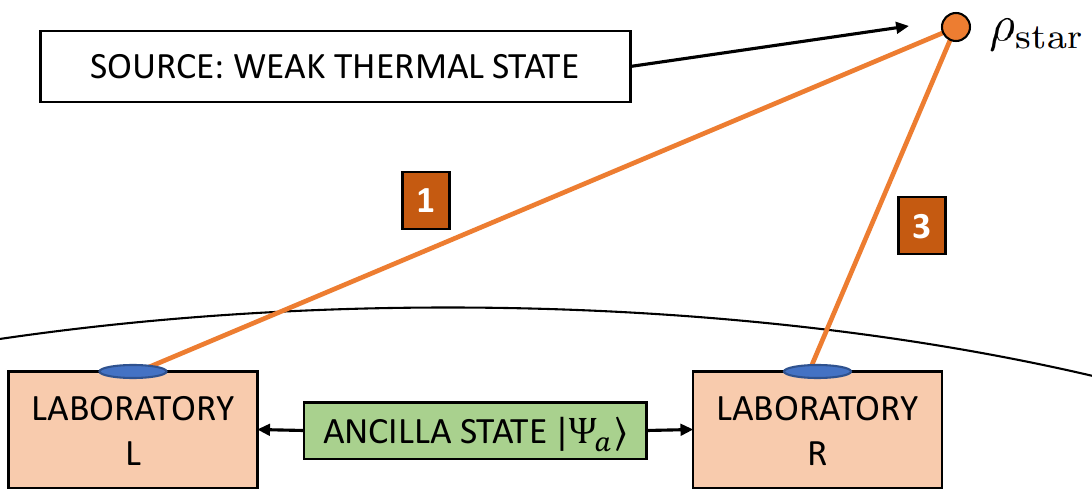}
    \caption{Generalized setup for quantum-assisted telescopy. Modes $1$ and $3$ couple the starlight to the left and right laboratories, respectively.}
    \label{fig:generalized_setup}
\end{figure}

\section{Setup}

To explain the basic principle of our procedures, we will consider the case where there are two telescopes that can receive the stellar photons. For weak sources, the average photon number per mode $\epsilon$ is much less than one, so we model the source as a weak thermal state~\cite{Tsang2011}, \be \label{eq:psi-star}
\begin{aligned}
    \rho_\text{star} = & (1-\epsilon)\ket{0_L 0_R}\bra{0_L 0_R} \\
         +  & \frac{\epsilon}{2}( \ket{1_L 0_R} \bra{1_L 0_R} + \ket{0_L 1_R} \bra{0_L 1_R} \\
        + & \mathcal{V}^* \ket{1_L 0_R}\bra{0_L 1_R} +\mathcal{V} \ket{0_L 1_R}\bra{1_L 0_R}) 
        + \mathcal{O}\left( \epsilon^2 \right), 
\end{aligned}
\ee
where $\ket{1_L 0_R}$ corresponds to one photon coming to the left ($L$) telescope and zero photons coming to the right ($R$) telescope, and similar for the other terms. The state \eqref{eq:psi-star} represents a theoretical model of incoming stellar radiation, where one assumes that the probability of getting more than one stellar photon at the telescopes is negligible.
% Based on \cite{mandel1961photon_degeneracy}, we perform a qualitative analysis of this assumption in the Supplementary Material.

We assume that $\epsilon$ is a known, small 
\footnote{The assumption of small $\epsilon$ is reasonable since the number of photons arriving from a black body source in the same quantum state at a given time is given by Planck distribution. For optical frequencies, this number is very much smaller than unity for all sources that are not much hotter than a few thousand degrees, because their peak emissivity lies in the infrared \cite{mandel1979photon}.}
parameter that can be estimated by other methods that are less sensitive to noise \cite{Tsang2011, J_Monnier_review}. Observe that if Eq. (\ref{eq:psi-star}) is valid, but the telescopes receive photons from sources other than the one of interest, then the information about these sources will be encoded within the visibility function. For example, the constant background, when Fourier transformed, becomes a sharp peak near the origin.

The goal of the measurement scheme is often supported by an ancilla state that is interacted with the incoming stellar state (\ref{eq:psi-star}). The goal is to extract information about the visibility $\mathcal{V}$ from the incoming state. In the end, the probabilities of possible measurement outcomes should depend on $\mathcal{V}$. 
%The interaction between the state (\ref{eq:psi-star}) and the ancilla resembles adding the signals in classical interferometric techniques, but the fact that one operates with quantum states rather than classical signals must be taken into account as one compares the classical and quantum techniques. For example, the discussion on shot noise, a major source of noise in classical interferometry \cite{bachor1989quantum}, becomes more subtle when we extend it to the measurements on quantum states, as discussed in more detail in the Supplementary Material.

One can treat each mode as a single-rail optical qubit where the computational basis states are the absence (0) and presence (1) of a photon. The protocol's goal is to determine the \textit{complex visibility} $\mathcal{V}$, which depends on the source intensity distribution and is a function of the baseline connecting two telescopes. Given the visibility as a function of baseline, one can use the van Cittert--Zernike theorem to determine the intensity profile of the source \cite{van_Cittert_Zernike_1, van_Cittert_Zernike}. For all protocols, we will consider both a point source, for which $\mathcal{V}=e^{-i\Phi}$, as well as an extended source, for which $\mathcal{V}=ge^{-i\Phi}$, where $g$ is a real and positive amplitude.

The star photons can arrive at two distant telescopes with separate laboratories at their locations, as shown in Fig.~\ref{fig:generalized_setup}. The laboratories share a known ancilla quantum state $\ket{\Psi_\text{a}}$ as a resource. We allow local operations and measurements within each laboratory, as well as classical communication between the laboratories, but we do not allow the distribution of stellar photons between the two laboratories. By definition, this prevents the loss of stellar photons that otherwise occurs during transmission over long baselines, and which limits existing long-baseline interferometry methods.

\section{Fisher Information}
To quantify and compare the information obtained by specific measurement schemes, we use the Fisher information $f(\Phi,g)$. It quantifies the information one obtains about the parameters to be estimated per measurement act. According to the Cramér-Rao bound, the inverse of the Fisher information matrix sets a lower bound on the covariance matrix describing the phase and amplitude estimation problem~\cite{cramer1946mathematical,rao1992information}. The upper bound on the Fisher information (FI) of a quantum measurement on the stellar photon state \eqref{eq:psi-star},
optimized over all possible measurement schemes, is given by the quantum Fisher information (QFI). 

The Fisher information (FI) matrix is given by
\begin{equation} \label{eq:fisher2} 
    f(\Phi,g) = \sum_{k} \frac{1}{p_k}
    \begin{pmatrix}
        \left( \frac{\partial p_k}{\partial \Phi} \right)^2 & \frac{\partial p_k}{\partial \Phi}\frac{\partial p_k}{\partial g} \\
        \frac{\partial p_k}{\partial g} \frac{\partial p_k}{\partial \Phi} & \left( \frac{\partial p_k}{\partial g} \right)^2
    \end{pmatrix},
\end{equation}
where $p_k$ is the probability of obtaining measurement outcome $k$. $f(\Phi,g)$ quantifies the information one obtains about the parameters to be estimated per measurement act. 

It quantifies the sensitivity of a given measurement scheme, and it saturates quantum Fisher information (QFI) if the measurement scheme is optimal, i.e., if the scheme extracts the most possible information about the estimated parameter.
The QFI has matrix elements
\begin{equation}
    h_{ij}(\Phi,g) = \text{Tr}\left[ \rho_\text{star} \frac{L_i L_j + L_j L_i}{2} \right], \ i,j \in \{ g, \Phi \},
\end{equation}
where $L_i$ is the symmetric logarithmic derivative (SLD) corresponding to parameter $i$, defined by
\begin{equation} \label{eq:SLD_def2}
    \frac{L_i \rho_\text{star} + \rho_\text{star} L_i}{2} = \partial_i \rho_\text{star}.
\end{equation}
This relation is satisfied by 
\begin{equation}
\begin{aligned}
    L_\Phi &= ig
    \begin{pmatrix}
        0 & -e^{-i\Phi} \\
        e^{i\Phi} & 0
    \end{pmatrix} \\
    L_g &= \frac{1}{1-g^2}
    \begin{pmatrix}
        -g & e^{-i\Phi} \\
        e^{i\Phi} & -g
    \end{pmatrix},
\end{aligned}
\end{equation}
using the basis $\ket{1_L 0_R}$ and $\ket{0_L 1_R}$. 

If one forbids physically bringing the stellar modes together (e.g., due to loss associated with long baselines), then the quantum-enhanced schemes offer an advantage over the ones known from classical telescopy. As shown in \cite{Tsang2011}, for any scheme without the resource of shared entanglement between the telescopes, the FI scales with $\epsilon^2$. Such scaling is achieved in heterodyne and homodyne detection. However, for state (\ref{eq:psi-star}), QFI scales with $\epsilon$, not $\epsilon^2$; FI can achieve such scaling only if entangled states are available.

We define any protocol whose FI saturates the QFI, $f(\Phi,g)=h(\Phi,g)$, as an optimal measurement scheme. It is not always possible to saturate this bound in the multi parameter case, as is the case for this particular system since the SLD matrices do not commute on the support of $\rho_\text{star}$~\cite{yang2019optimal}. Therefore, we will focus on estimating the phase $\Phi$ (by setting $g=1$). We will present two protocols that are optimal for this single-parameter case. We will denote the single-parameter FI and QFI by $f(\Phi)$ and $h(\Phi)$.

The pioneering nonlocal scheme of Gottesman \textit{et al.} presented in \cite{Gottesman2012} has a FI of $f(\Phi)=h(\Phi)/2$, so although it has certain advantages over classical interferometry, it is not an optimal scheme. This result reflects the fact that only half the star photons are used for the estimation in that particular scheme. We propose a protocol that uses all the star photons and thus gives twice the precision in the estimate of $\Phi$. The improvement of FI is also achieved in the two-parameter case where $\mathcal{V}=ge^{-i\Phi}$. However, this improvement does not make the protocol saturate QFI, as should be expected for the reasons described before.

\section{NOT-based protocol}

The protocol of Gottesman \textit{et al.}~\cite{Gottesman2012} uses only linear optical elements to achieve half of the quantum Fisher information. We show in Appendix~\ref{app:linear_optics} that this is in fact the best one can get with the ancilla from their proposal and linear optical elements. To achieve an optimal measurement scheme, we propose the use of nonlinear components. In this case, we make use of an optical NOT gate in the Fock basis. That is, performing the gate results in flipping the state of the mode, i.e., a photon in the mode is either created if there is no incoming photon, or destroyed if there is a single incoming photon. We emphasize that in this paper we assume the existence of such a gate which is allowed by quantum mechanics, but at the time of writing this manuscript there is no known theoretical proposal or experimental realization of it.

\begin{figure}
    \centering
    \includegraphics[width=8.6cm]{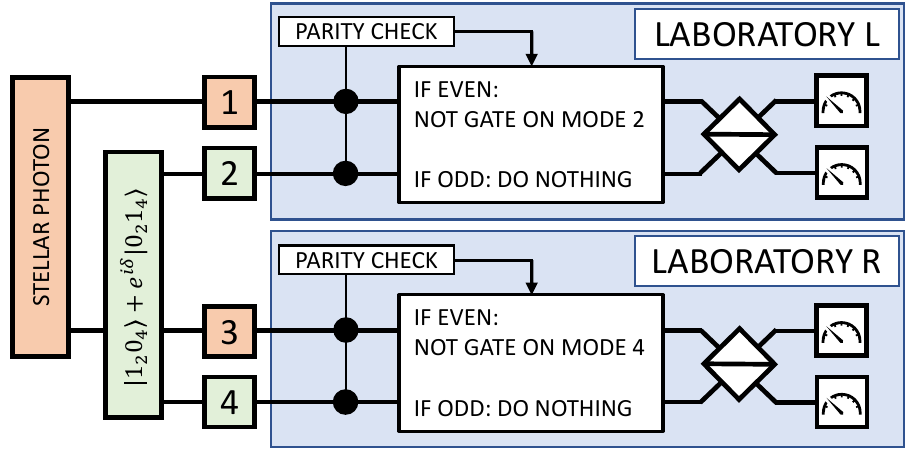}
    \caption{Circuit representation of the nonlocal phase estimation protocol. We assume a phase shift of $i$ upon reflection at the beam splitter. }
    \label{fig:nonlocal_setup}
\end{figure}

We consider a four-mode protocol and provide both laboratories with the ancilla state
\be \label{eq:ancilla_nonlocal_protocol}
\ket{\Psi_\text{a}} = 
\frac{1}{\sqrt{2}}\big( \ket{1_2 0_4} + e^{i\delta}\ket{0_2 1_4}),
\ee 
where the modes $2$ and $4$ are associated with a single-photon entangled state, and $\delta$ is a tunable phase. The subscripts indicate the modes indicated in Fig. \ref{fig:nonlocal_setup}; modes $1$ and $2$ reach laboratory $L$, modes $2$ and $4$ reach laboratory $R$. The total state received by both laboratories is then $\rho_0 = \rho_\text{star} \otimes \ket{\Psi_\text{a}} \bra{\Psi_\text{a}}$. 

The state $\rho_0$ undergoes the series of operations shown in Fig.~\ref{fig:nonlocal_setup}, and is measured in the number basis. The parity checks verify whether the total number of photons within a single laboratory is even or odd. If they return the same result, then the stellar photon has arrived, otherwise it has not (see Appendix~\ref{app:parity-calculation} for more detailed calculations). If both laboratories obtained an \textit{even} ($e$) result, the probabilities of possible outcomes are
\begin{equation} \label{eq:prob11}
\begin{aligned}
    & p(1_1 0_2 1_3 0_4, e,e) = p(0_1 1_2 0_3 1_4, e,e)     =\frac{\epsilon}{8}\left[1 - \cos (\Phi+\delta) \right] \\
    & p(1_1 0_2 0_3 1_4, e,e) = p(0_1 1_2 1_3 0_4, e,e)  = \frac{\epsilon}{8}\left[1 + \cos (\Phi+\delta) \right] 
\end{aligned}
\end{equation}
and if both the parity check results were \textit{odd} ($o$)

\begin{equation} \label{eq:prob12}
\begin{aligned}
    & p(1_1 0_2 1_3 0_4,o,o) = p(0_1 1_2 0_3 1_4,o,o) = \frac{\epsilon}{8}\left[1 + \cos (\Phi-\delta) \right] \\
    & p(1_1 0_2 0_3 1_4,o,o) = p(0_1 1_2 1_3 0_4,o,o)  = \frac{\epsilon}{8}\left[1 - \cos (\Phi-\delta) \right].
\end{aligned}
\end{equation}
The results for which the parity measurements do not agree should be discarded, as they correspond to the cases where no stellar photon arrived. Each parity check can be implemented with a pair of CNOT gates and an ancillary qubit (a more detailed explanation is given in Appendix~\ref{app:parity-calculation}). For an extended source, one replaces $\cos(\Phi\pm\delta)\rightarrow \text{Re}\left\{ \mathcal{V} e^{\mp i \delta} \right\}$. 

Equations (\ref{eq:prob11}) and (\ref{eq:prob12}) then allow one to estimate the relative phase shift $\Phi$. Classical communication between the laboratories is required to determine which coincidence occurred after all the measurements are performed.

We can determine whether the protocol described in this section is optimal by evaluating the FI $f$ using (\ref{eq:prob11}) and (\ref{eq:prob12}) and then comparing it to the QFI. For the phase-estimation problem, the resulting FI is $f(\Phi) = \epsilon$, which saturates QFI. It shows that this protocol is an optimal measurement of the relative phase shift $\Phi$, and gains twice as much information per stellar photon as the protocol using only linear elements. A similar improvement is also achieved in the case of single-parameter estimation of the visibility amplitude $g$ (more details in Appendix \ref{app:fisher}). Finally, for the complex visibility $\mathcal{V}$ estimation problem, our protocol achieves an improvement in FI over the Gottesman \textit{et al.} procedure, which is quantified in Appendix \ref{app:fisher}; however, the QFI is not saturated.

%In reality, the stellar photon is in a weak thermal state~\cite{Tsang2011}, so there is a large probability that no photon arrives at either telescope. Crucially, one can determine whether or not a photon arrived by comparing the measurement results of qubits $0$ and $5$. If they are the same, then a stellar photon arrived; if they are different, then no stellar photon arrived. 

A possible error is the loss of the entangled ancilla corresponding to $\ket{0_2 0_4}$ in the input. Such an error cannot be identified by a single detection event since it leads to a set of results similar to the one corresponding to the procedure without error, However, it can be identified by examining the frequency of the ($0_0 0_5$) and ($1_0 1_5$) events: if the error is introduced, the latter events occur more often. This detection scheme works only if the error appears on a recurrent basis. We perform a more detailed error analysis in Appendix \ref{app:parity-calculation}. 

The NOT-based protocol offers an improvement over the proposal of Gottesman \textit{et al.}, but it requires NOT quantum gates and the ability to perform a parity check for optimal performance. Implementing the protocol optically would require deterministic nonlinear optical gates, beyond what is currently available. Moreover, the nonlinear NOT gates and the parity checks need to be extremely reliable, otherwise the errors dominate the signal used for the visibility estimation. The errors can be reduced if one introduces additional elements that verify the arrival of the stellar photon without destroying the visibility information. More details are provided in Appendix \ref{app:gate-errors}. Even though our proposal makes use of elements that are beyond what is currently feasible, we hope it will stimulate research in developing the required NOT gates and parity checks, as they would also have applications beyond our proposal. Encouraged by the rapid development of quantum information processing both in theory and experiments, we hope that our proposal will be possible to achieve in the future.

In the event that the optical setup remains infeasible, another approach (motivated by recent developments in quantum transduction \cite{mirhosseini2020superconducting,transduction_overview}) is to use nonphotonic ancilla qubits that are easier to manipulate, and to transduce them into photonic qubits before the beam splitters in Fig. \ref{fig:nonlocal_setup}.

The protocol of Gottesman \textit{et al.} can be used as a reference to determine the validity of using the \textit{NOT-based protocol}. For the latter protocol, observe that if one removes the parity check and the NOT gates, one recovers the original protocol of Gottesman \textit{et al.} However, introducing new elements to the circuit can introduce new types of errors. The \textit{NOT-based protocol} can be considered as a valid improvement only if the errors are small enough that one can extract more information about the visibility from each stellar photon when compared to the Gottesman \textit{et al.} protocol. We analyze the influence of errors due to the new elements in Appendix \ref{app:gate-errors}.

\section{Modified quantum memory protocol}
Even though our NOT-based protocol is an optimal phase measurement scheme, it requires a copy of the ancilla state for each possible time bin (more precisely, for each possible temporal mode within the duration of the measurement and over the bandwidth of the collected starlight); this requires a large amount of resources and is experimentally infeasible. Khabiboulline \textit{et al.} \cite{Khabiboulline} proposed an optimal phase measurement scheme that encodes the arrival time of the star photon in a quantum memory, for which the amount of required resources scales logarithmically with the number of time bins. We propose a modification to their scheme that both simplifies it and reduces the required resources by half, which is potentially critical for the practical implementation of these ideas.

Consider the modes provided by the star as single-rail qubits, where the logical $0$ and $1$ denote the absence or presence of a single photon in a mode. Suppose we can measure them in an arbitrary basis. If we know the star provided a photon, then the optimal phase measurement is achieved when we directly measure both stellar modes. One measurement is done in the \textit{X} basis, spanned by $\ket{\pm} = \frac{1}{\sqrt{2}}(\ket{0}\pm\ket{1})$, and the other mode is measured in the rotated basis spanned by $\ket{\pm_\delta} = \frac{1}{\sqrt{2}}(\ket{0}\pm e^{i\delta}\ket{1})$. Given the setup in Fig.~\ref{fig:generalized_setup}, performing the \textit{X} basis measurement on mode $1$ and rotated basis measurement on mode $3$ results in the probabilities conditioned on the stellar photon arrival,
\begin{equation} \label{eq:kh_prob}
\begin{aligned}
    P(+,+_\delta) = P(-,-_\delta) = & \frac{1}{4} \left[1 + \cos(\Phi + \delta) \right] \\
    P(+,-_\delta) = P(-,+_\delta) = & \frac{1}{4} \left[1 - \cos(\Phi + \delta) \right].
\end{aligned}
\end{equation}
where $P(\pm,\pm_\delta)$ indicates the probability of result $\pm$ in laboratory $L$ and $\pm_\delta$ in laboratory R.
%Such measurements can be done on single-rail qubits in a non-deterministic and heralded way \cite{lund2002nondeterministic}, but the protocol will suffer from a lower Fisher information. Another solution could be transducing the single-rail optical qubit to another type of qubit for which the measurement would be easier.

The Fisher information for this set of probabilities saturates the QFI, so this is also an optimal phase measurement scheme. For extended sources, one replaces $\cos(\Phi+\delta)\rightarrow \text{Re}\left\{ \mathcal{V} e^{- i \delta} \right\}$. 

\begin{figure}
    \centering
    \includegraphics[width=8.6cm]{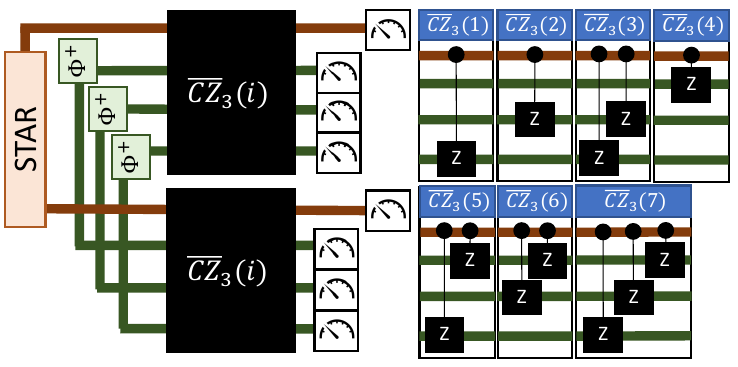}
    \caption{Scheme of the modified version of Khabiboulline's protocol for $N=7$ time bins. $\overline{\text{CZ}}_k(i)$ indicates a sequence of controlled phase gates with $k$ target qubits (ancilla) corresponding to the $i$th time bin.}
    \label{fig:modified_kh}
\end{figure}

The measurement on the stellar photon has an issue: we cannot tell if the star provided a photon, since the lack of arrival of the stellar photon can lead to the same results as in (\ref{eq:kh_prob}). We need to know whether or not the photon has arrived and, if it has, then we must know when it happened. This is achieved by the procedure shown in Fig.~\ref{fig:modified_kh}.

Suppose that within time $T$, we expect at most one photon to arrive from the star. We divide $T$ into $N$ short time bins of length $\tau$, corresponding to temporal modes, so that $T=N\tau$. To perform binary encoding of the time bin, we need $2\log_2(N-1)$ ancilla qubits, each prepared in the state $\ket{\Phi^+}$, where $\ket{\Phi^\pm}=(\ket{00}\pm\ket{11})/\sqrt{2}$ are maximally entangled Bell states. The first qubit from each pair is distributed to laboratory $L$, and the second qubit to $R$. The next step is to pass the ancilla through a set of controlled phase gates ($\text{CZ}$) that depends on the time bin, where 

\begin{equation}
\begin{aligned}
    \ket{0_c 0_t} \xrightarrow{CZ} \ket{0_c 0_t}, \quad & \ket{0_c 1_t} \xrightarrow{CZ} \ket{0_c 1_t}, \\
    \ket{1_c 0_t} \xrightarrow{CZ} \ket{1_c 0_t}, \quad & 
    \ket{1_c 1_t} \xrightarrow{CZ} -\ket{1_c 1_t}
\end{aligned}
\end{equation}
performs a standard phase shift gate $\text{Z}$ on a target qubit when the state of the control qubit is $1$. The index $c$ denotes the control qubit, and $t$ denotes the target qubit. A $\text{Z}$ gate acting on one of the qubits in a Bell pair can be used to switch between Bell states, $\text{Z}\ket{\Phi^\pm}=\ket{\Phi^\mp}$. In our case, the star supplies the control qubits for the gates and the ancilla supplies the target qubits. If the star photon arrives during the $n$th time bin, then the sequence of gates $\bigotimes_{i=1}^{2\log_2(N-1)}\text{Z}^{n_i}$ is performed on the locally available ancilla qubits, where $n_i$ is the $i$th digit of the integer $n$ written in binary (see Appendices \ref{app:memory_example} and \ref{app:memory_example_mod} for explicit examples).

This encodes the time bin information into the Bell states. A similar process was used in \cite{Khabiboulline}, but using an extra set of intermediary memory qubits which are the targets of a similarly modified CNOT gate before encoding the time bin information into the Bell states.

Similarly to the protocol of Khabiboulline \textit{et al.} \cite{Khabiboulline}, this protocol achieves a significant improvement in the entanglement cost when compared with the unary protocols. If within $N$ time bins one expects to get at most one stellar photon, then one needs $\lceil\log_2(N+1)\rceil$ entangled qubit pairs to perform the protocol given in this section, where $\lceil...\rceil$ is the ceiling function. On the other hand, unary protocols require $N$ entangled qubit pairs. An example of resource costs for the modified protocol is given in Appendix~\ref{app:resources}.

However, our modified protocol reduces the number of ancilla qubits and gates by half when compared to the proposal given in \cite{Khabiboulline}, by eliminating the intermediary memory qubits. This advantage can be achieved only if both parties are capable of performing the measurements of the stellar photon modes for all time bins and storing the results classically. It is necessary since parties must know when the stellar photon has arrived prior to the visibility measurement. Only then are they capable of selecting the result that is useful in the visibility estimation. We note that the scheme requires classical (not quantum) memory.

Finally, the protocol described in this section can serve as a subroutine for other unary protocols if performed for a single time bin. In Appendix \ref{app:gate-errors} we consider the advantages of merging the \textit{modified quantum memory protocol} and the \textit{NOT-based protocol.} 

\section{Conclusions}
We have proposed two quantum-enhanced long-baseline interferometry schemes that offer improvements over two prior proposals. The Gottesman \textit{et al.} protocol \cite{Gottesman2012} cannot be improved if one is limited by the ancilla, linear optics, and measurements in the photon number basis, but the development of nonlinear photonic gates or quantum transducers enables us to improve it and achieve an optimal phase-estimation scheme. Such a protocol achieves the maximum allowed value of Fisher information, but (similar to the Gottesman \textit{et al.} proposal) it consumes one copy of the ancilla state for each time bin. This linear scaling of resources was improved to logarithmic by Khabiboulline \textit{et al.} by using binary encoding to store the time of arrival of the stellar photon. We have modified their scheme to reduce the number of ancilla qubits and gate operations by half. This is done by encoding the time bin information directly into the Bell-state ancilla qubits, using controlled phase gates instead of using intermediary memory qubits with controlled NOT gates. 

\begin{acknowledgments}
We thank Eric Chitambar, Virginia O. Lorenz, John D. Monnier, Michael G. Raymer, and Brian J. Smith for helpful discussions. This work was supported by the multi-university National Science Foundation Grant No. 193632, QII-TAQS: Quantum-Enhanced Telescopy.
\end{acknowledgments}

\appendix
\section{Gate errors in the NOT-based protocol} \label{app:gate-errors}

We consider the errors introduced by the gates in the NOT-based protocol with the assumption that other elements of the circuit (beam splitters, detectors, distribution of the entangled ancilla) work ideally. Our goal is to analyze the influence of elements of our circuit that are not present in the protocol of Gottesman \textit{et al.} The analysis will be performed for the scheme depicted in Fig. 2 in the main body of the paper.

Denote by $p$ the probability that the parity check is performed correctly, and by $q$ the probability of performing the NOT gate without an error. We assume that with probability $(1-q)$, the gate leaves the state unmodified.

The setup can result in several different scenarios, which can be labeled by the parity of the incoming global state and various combinations of the parity checks and NOT gates working correctly or failing. For a stellar point source, the incoming global state given assuming the stellar photon has arrived is
\begin{equation}
\begin{aligned}
     \ket{\psi_0} & = \frac{1}{\sqrt{2}}\left( e^{i\Phi}\ket{1_1 0_3} + \ket{0_1 1_3}\right) \\
     & \otimes \frac{1}{\sqrt{2}}
    (\ket{1_2 0_4} + e^{i\delta}\ket{0_2 1_4})   \\
    & = \frac{1}{2}\big( e^{i\Phi} \ket{1_1 1_2 0_3 0_4} + e^{i\delta} \ket{0_1 0_2 1_3 1_4} \\
    & + e^{i(\Phi+\delta)}\ket{1_1 0_2 0_3 1_4} + \ket{0_1 1_2 1_3 0_4}\big)
\end{aligned}
\end{equation}

After the final equal sign, the first line corresponds to the correct parity results of \textit{(even, even)} and the second line to \textit{(odd, odd)}. The probability of receiving an \textit{(even, even)} state and performing all the parity checks and NOT gates successfully is $(\epsilon/2)p^2 q^2$. The probability of receiving an \textit{(odd, odd)} state and performing the parity checks successfully is $(\epsilon/2)p^2$, where the $q^2$ factor is missing since the NOT gates are not performed when the local parity result is \textit{odd}. The probability of receiving the stellar photon and performing all the local operations correctly on it is
\begin{equation}
    P_+ = \frac{\epsilon}{2}p^2q^2 + \frac{\epsilon}{2}p^2 = \frac{\epsilon}{2} p^2 (1+q^2).
\end{equation}
This corresponds to the fraction of events where the information about the stellar phase $\phi$ is extracted correctly. 

If the stellar photon has not arrived, then the incoming stellar state is $\ket{0_1 0_3}$ resulting in the global state
\begin{equation}
    \ket{\psi_0} = \frac{1}{\sqrt{2}}(\ket{0_1 1_2 0_3 0_4} + e^{i\delta}\ket{0_1 0_2 0_3 1_4}).
\end{equation}
The kets in the above equation correspond to the parity measurement of \textit{(odd, even)} and \textit{(even, odd)}. 

Caution must be taken when performing the protocol since there are possible scenarios in which the errors in the parity checks and NOT gates lead to output measurement results included in Eq. (\ref{eq:prob11}), corrupting the result of that equation. The first scenario when it can happen corresponds to the input parity state (\textit{odd}, \textit{odd}) and performing all parity checks and NOT gates incorrectly. This happens with the probability $(\epsilon/2)(1-p)^2(1-q)^2$.

The second fraudulent scenario occurs for the incoming parity state of (\textit{even}, \textit{odd}), performing the parity measurement in the right laboratory incorrectly (the measured parities are \textit{even, even}), and failing to perform the NOT gate in the right laboratory. The probability of such an event is $(1-\epsilon)p(1-p)q(1-q)$. The third fraudulent scenario is symmetric to the second one, but for the incoming parity state of (\textit{odd}, \textit{even}). 

The probability of confusing a fraudulent event for a correct one is 
\begin{equation} \label{eq:Pf}
    P_f =  \frac{\epsilon}{2}(1-p)^2(1-q)^2 + (1-\epsilon)p(1-p)q(1-q).
\end{equation}

The ratio of the events valid for the visibility estimation to the fraudulent events is
\begin{equation}
    \frac{P_+}{P_f} = \cfrac{\frac{\epsilon}{2} \  p^2 (1+q^2)}{\frac{\epsilon}{2}(1-p)^2(1-q)^2 + (1-\epsilon)p(1-p)q(1-q)},
\end{equation}
and it goes to zero in the weak thermal light regime $\epsilon\rightarrow 0$. For reasonable values of $p$ and $q$, the protocol is highly susceptible to errors since the overwhelming majority of the events included in the visibility estimation is fraudulent. 

\begin{figure*}
    \centering
    \includegraphics[width=0.8\textwidth]{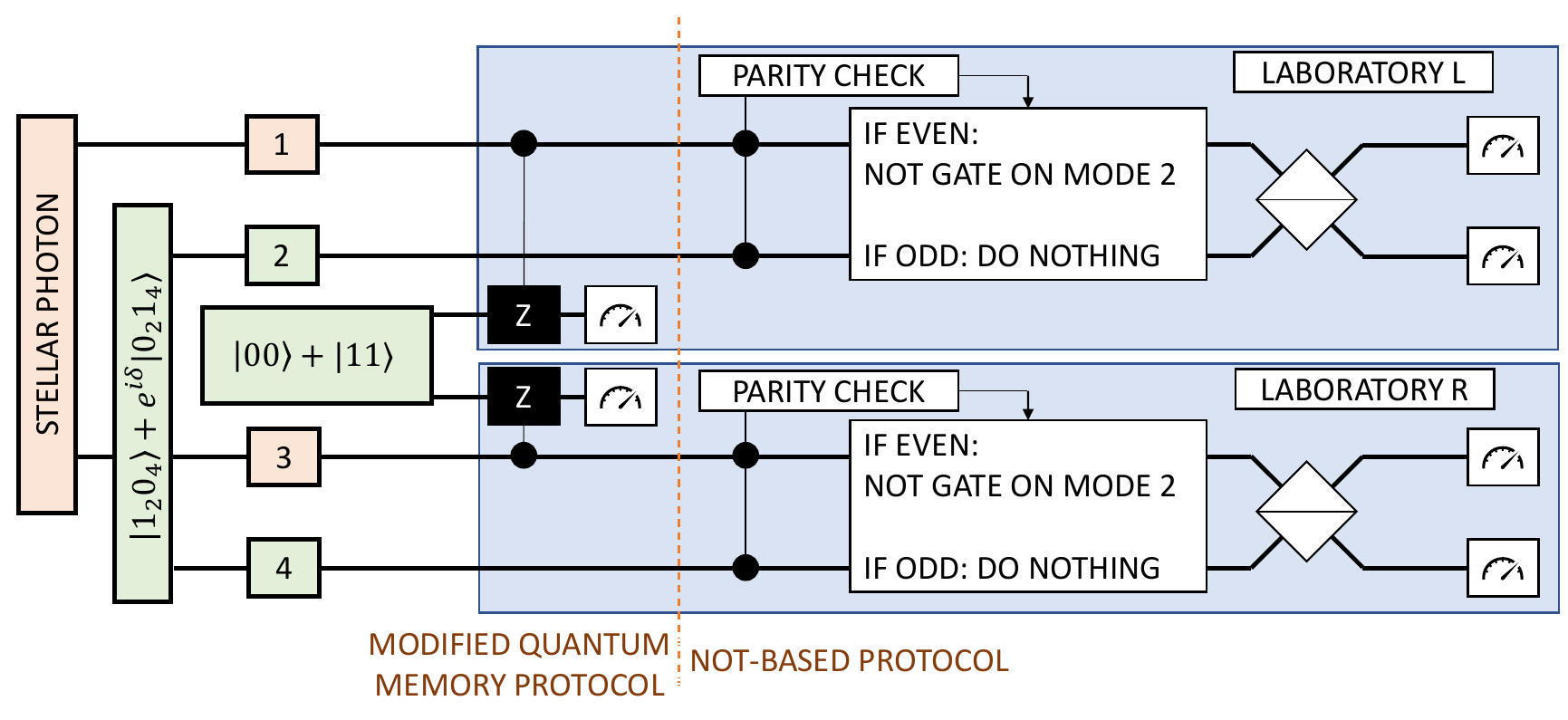}
    \caption{Scheme of the \textit{NOT-based protocol} with the \textit{modified quantum memory protocol} included as a subroutine.}
    \label{fig:merged}
\end{figure*}

The protocol can be improved if one includes an efficient measurement verifying the arrival of the stellar photon. 
This can be done by merging the \textit{NOT-based protocol } with the \textit{modified quantum memory protocol} performed for one time bin at the cost of distributing an additional entangled Bell pair per time bin, as shown in Fig. \ref{fig:merged}. It allows one to postselect on the stellar photon arrival events, and therefore discard the fraudulent events that can be potentially confused for the ones that are valid for the visibility estimation. Such fraudulent events correspond to the second term in (\ref{eq:Pf}). As discussed in the previous section, the \textit{modified quantum memory protocol} requires performing the $CZ$ gates; we will denote by $\eta$ the probability that such gate works correctly and by $(1-\eta)$ the probability that the gate leaves the state unchanged.

\begin{figure}
    \centering
    \includegraphics[width=8.6cm]{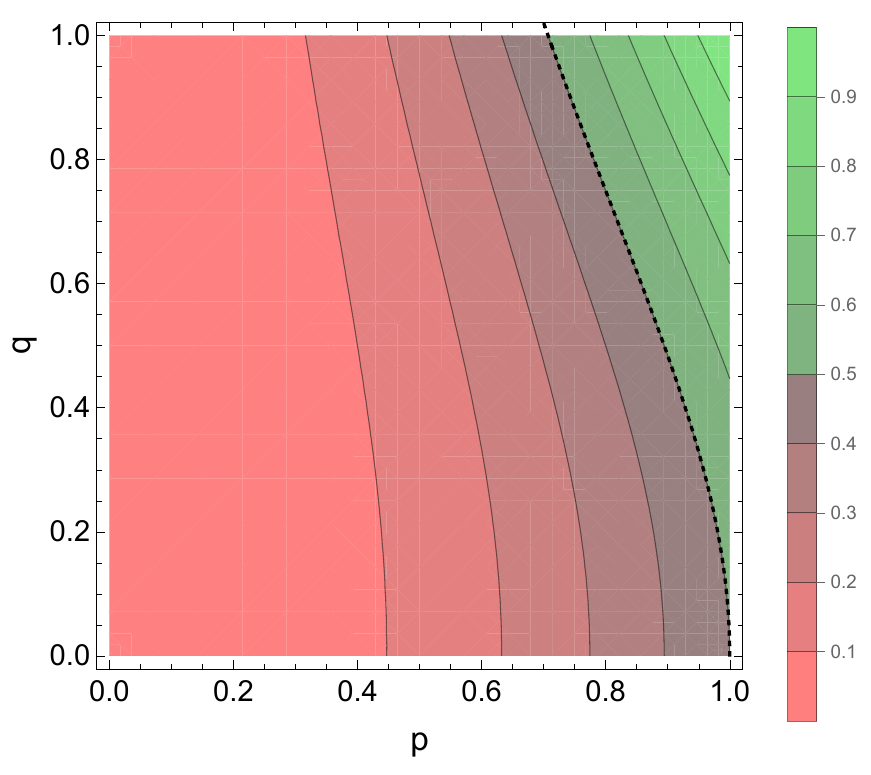}
    \caption{Contour plot of $P_+'$ defined by Eq. (\ref{eq:pPlus}) for $\eta=1$. The protocol of Gottesman \textit{et al.} is outperformed only in the region of the plot for which the values exceed $0.5$, as indicated by the green color. The performance of the Gottesman \textit{et al.} protocol is matched along the dotted line. For $\eta<1$, all the values on the plot are reduced by a factor of $\eta$.}
    \label{fig:pPlus}
\end{figure}

\begin{figure}
    \centering
    \includegraphics[width=8.6cm]{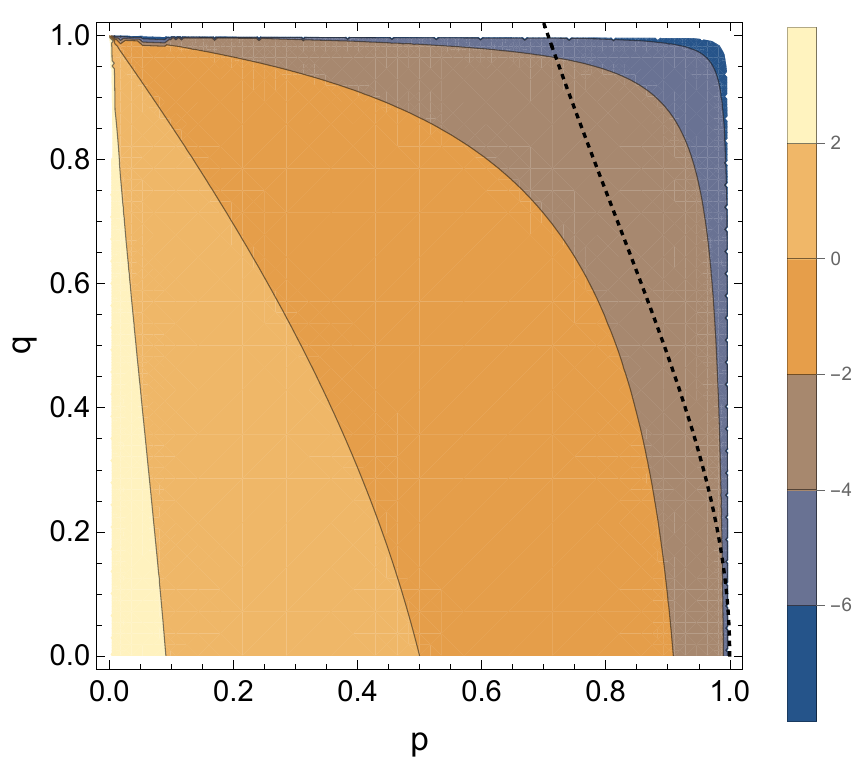}
    \caption{Contour plot of $\log_{10}[P_f' / (P_+'+P_f')]$, with $P_+'$ and $P_f'$ defined in (\ref{eq:pPlus}) and (\ref{eq:pF}). The Gottesman \textit{et al.} protocol can be outperformed for $p$ and $q$ values to the right of the dotted line.}
    \label{fig:ratio}
\end{figure}

If one postselects the events when the stellar photon has arrived, then the probability of including the proper events for the visibility estimation becomes
\begin{equation} \label{eq:pPlus}
    P_+' = \frac{\eta p^2 (1+q^2)}{2} 
\end{equation}
and the probability of confusing a fraudulent event for a valid one is 
\begin{equation} \label{eq:pF}
    P_f' = \eta (1-p)^2(1-q)^2/2.
\end{equation}
The new scheme outperforms the original Gottesman protocol if one uses more than half of the stellar photons to determine the visibility; the possible values of $p$ and $q$ for which this is possible are indicated in Fig. \ref{fig:pPlus}. For these values, the ratio of the fraudulent events to the events that are valid for visibility estimation is negligible (see Fig. \ref{fig:ratio}). 

\section{Gottesman \textit{et al.} Protocol - Limitations of the Ancilla and Linear Optics} \label{app:linear_optics}

\begin{figure} 
    \centering
    \includegraphics[width=8.6cm]{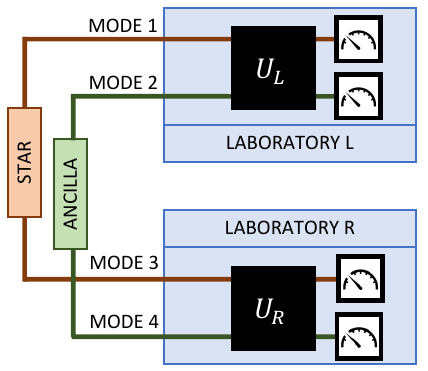}
    \caption{Scheme of generalized Gottesman \textit{et al.} protocol. The ancilla and measurements are kept the same as in the original proposal. We restrict this analysis to the local operations $U_L$ and $U_R$.}
    \label{fig:gottesman_analysis}
\end{figure}

Consider the setup given in Fig.~\ref{fig:gottesman_analysis}. Modes 1 and 3 are supplied by the star, and modes 2 and 4 are the ancilla given by
\begin{equation} \label{eq:gottesman_ancilla_2}
    \ket{\Psi_\text{a}} = \frac{1}{\sqrt{2}}(\ket{1_2 0_4} + e^{i\delta}\ket{0_2 1_4}),
\end{equation}
where the indices indicate the relevant modes. We require that the measurements are performed in the photon number basis, and that the manipulations of modes are local, i.e., the black boxes evolve the input state according to a unitary operation $U=U_L \otimes U_R$. Hrer, $U_L$ acts only on modes 1 and 2, and $U_R$ acts on 3 and 4. Assume that $U_L$ and $U_R$ represent sets of linear optical elements that do not change the local number of photons, but are otherwise arbitrary. 

For simplicity, we will take the star to be a point source that either supplies the vacuum or a single photon, in the state
\begin{equation}
    \rho = (1-\epsilon)\ket{0_L 0_R}\bra{0_L 0_R} + \epsilon \ket{\Psi_1}\bra{\Psi_1},
\end{equation}
where
\be 
\ket{\Psi_1} = \frac{1}{\sqrt{2}}\left( e^{i\Phi}\ket{1_1 0_3} + \ket{0_1 1_3}  \right).
\ee
It is possible to filter out the vacuum events since in those cases the two meters will detect exactly one total excitation, which comes from the ancilla, since the protocol preserves photon number. For the $\ket{\Psi_1}$ events, we can take the input state of the circuit to be
\begin{equation} \label{eq:input_gottesman}
\begin{aligned}
    \ket{\Psi_1}\otimes\ket{\Psi_\text{a}} 
    = \frac{1}{2}\Big( 
    & e^{i\Phi} \ket{1_1 1_2 0_3 0_4} + e^{i\delta}\ket{0_1 0_2 1_3 1_4} \\
    + & e^{i(\Phi+\delta)} \ket{1_1 0_2 0_3 1_4} + \ket{0_1 1_2 1_3 0_4} \Big).
\end{aligned}
\end{equation}
Applying the $U$ operator gives
\begin{equation} \label{eq:outupt_generalized_gottesman}
\begin{aligned}
    \ket{\Psi} = \frac{1}{2}\Big( 
    & e^{i\Phi} U_L\ket{1_1 1_2} \otimes \ket{0_3 0_4} + e^{i\delta}\ket{0_1 0_2} \otimes U_R\ket{1_3 1_4} \\
+   & e^{i(\Phi+\delta)} U_L\ket{1_1 0_2} \otimes U_R\ket{0_3 1_4} \\
+   & U_L\ket{0_1 1_2}     \otimes U_R\ket{1_3 0_4} \Big),
\end{aligned}
\end{equation}
where we assume that $U_{L}$ and $U_R$ leave the vacuum unchanged. Regardless of the choice of $U_{L}$ and $U_R$, the set of results due to the first and second terms will not overlap with the results due to the other terms because of the different numbers of photons. Since $e^{i\Phi}$ acts as a global phase shift for the first term, it will not affect the probabilities of the measurement results. Therefore, the first and second terms will not provide any information about $\Phi$, and we can postselect on the final two terms containing one photon. Note that it is possible to do so because the first two terms correspond to observing two excitations in one laboratory and no excitations in the other one.

The final two terms in (\ref{eq:outupt_generalized_gottesman}) can have $\Phi$-dependent interference fringes after performing $U_{L}$ and $U_R$. Evaluating the QFI from these terms results in $h=1$, which should be corrected for the fact that the probability of observing the corresponding events is $\epsilon/2$ [where $\epsilon$ is the probability that the star supplies a single photon to the receivers, and $1/2$ is the probability of observing an event due to the final two terms in (\ref{eq:input_gottesman}) given that a photon has arrived]. Therefore, the upper bound on the Fisher information, corresponding to the best choice of $U_{L}$ and $U_R$, is $\epsilon/2$. This is exactly the value obtained for the Gottesman \textit{et al.} protocol.

We conclude that the Gottesman \textit{et al.} protocol cannot be improved if we restrict ourselves to setups as in Fig.~\ref{fig:gottesman_analysis}, with an ancilla given by (\ref{eq:gottesman_ancilla_2}), using only linear optics and measurements in the number basis. Any improvement to the Gottesman protocol requires breaking one of these requirements in order to extract information about $\Phi$ from all the terms in (\ref{eq:outupt_generalized_gottesman}). In our first protocol (Fig. 2), we used nonlinear elements that enable us to perform the NOT gates. 
Even though our scheme offers improvements over the Gottesman \textit{et al.} protocol, it is a challenge to develop a physical operation that applies these gates.

\section{NOT-Based Protocol: Parity Check and Example Calculation} \label{app:parity-calculation}

\begin{figure} 
    \centering
    \includegraphics[width=8.6cm]{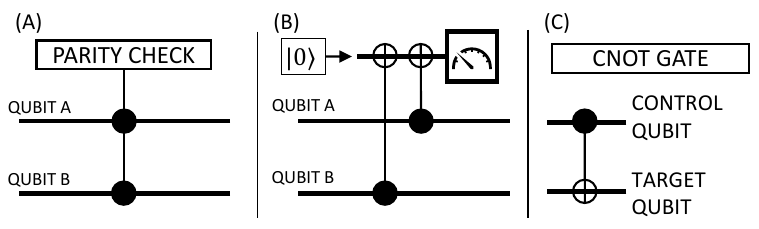}
    \caption{(A) Symbolic representation of the parity check. (B) Quantum circuit that performs the parity check. (C) Circuit representation of the CNOT gate.}
    \label{fig:equivalence}
\end{figure}

%Note: initial state of the ancilla used for parity checks different than in the paper
The parity check used within the NOT-based protocol can be performed with a pair of CNOT gates and an ancilla qubit, as indicated in Fig. \ref{fig:equivalence}. Suppose that we want to perform the parity check on qubits A and B by the setup given in Fig. \ref{fig:equivalence}B. We initialize the ancilla qubit in the state $\ket{0}$. If both A and B are in an even state ($\ket{0_A 0_B}$ or $\ket{1_A 1_B}$), then the state of the ancilla qubit remains unchanged after the CNOT gates. If they are in an odd state ($\ket{1_A 0_B}$ or $\ket{0_A 1_B}$), then the state of the ancilla qubit is flipped from $\ket{0}$ to $\ket{1}$. The measurement of the ancilla qubit in the number basis returns the parity of the qubits A and B. 

We will perform an example calculation for the NOT-based protocol by taking the star to be a point source that either sends a vacuum state or a single photon to the observer. The diagram of the setup is given in the \textit{NOT-based protocol} section in the main body of the paper. As we noted, the parity checks will require an additional ancilla qubit within each laboratory. We will denote these qubits by the indices $0$ (laboratory \textit{L}) and $5$ (laboratory \textit{R})

If the star sends the vacuum (modes $1$ and $3$ in state $\ket{0}$), then the input state is
\begin{equation}
    \frac{1}{\sqrt{2}}\left(
    \ket{0_0 0_1 1_2 0_3 0_4 0_5} + e^{i\delta}\ket{0_0 0_1 0_2 0_3 1_4 0_5}
    \right).
\end{equation}
Here we use mode $0$ to perform a parity check of modes $1$ and $2$, and mode $5$ for parity check of modes $3$ and $4$. After the CNOT gates performed within the parity check, the state becomes
\begin{equation}
    \frac{1}{\sqrt{2}}\left(
    \ket{1_0 0_1 1_2 0_3 0_4 0_5} + e^{i\delta}\ket{0_0 0_1 0_2 0_3 1_4 1_5}
    \right).
\end{equation}
Observe that within both kets, the states of modes $0$ and $5$ do not agree. Therefore, the parity measurement results performed within both laboratories will not agree if the star has not provided the photon. 

If, instead, the star supplies the photon, the input state is
\begin{equation} 
\begin{aligned}
     \frac{1}{2}
     & \Big( e^{i\Phi} \ket{0_0 1_1 1_2 0_3 0_4 0_5} 
        + e^{i\delta} \ket{0_0 0_1 0_2 1_3 1_4 0_5} \\
     & + e^{i(\Phi+\delta)} \ket{0_0 1_1 0_2 0_3 1_4 0_5} 
        + \ket{0_0 0_1 1_2 1_3 0_4 0_5} \Big).
\end{aligned}
\end{equation}
Performing the CNOT gates within the parity checks modifies that state to 
\begin{equation} 
\begin{aligned}
     \frac{1}{2}
     & \Big[ \big(e^{i\Phi} \ket{1_1 1_2 0_3 0_4} 
        + e^{i\delta} \ket{0_1 0_2 1_3 1_4}\big)\otimes\ket{0_0 0_5} \\
     & + \big(e^{i(\Phi+\delta)} \ket{1_1 0_2 0_3 1_4} 
        + \ket{0_1 1_2 1_3 0_4}\big)\otimes\ket{1_0 1_5} \Big].
\end{aligned}
\end{equation}
The states of the qubits $0$ and $5$ agree within all kets in the state above. Therefore, the parity check results should agree if the stellar photon has arrived. The next step is to perform the measurements of qubits $0$ and $5$ that return the parity results and postselect one of the states in the round $(\cdot)$ brackets. If the local parity measurement results returned \textit{even} ($\ket{0}_0$, $\ket{0}_5$), then the NOT gates are performed on the qubits $2$ and $4$. It results in the following state of modes $1$--$4$ after the parity checks:
\begin{equation}
    \frac{1}{\sqrt{2}}\left(
    e^{i(\Phi\pm\delta)}\ket{1_1 0_2 0_3 1_4} + \ket{0_1 1_2 1_3 0_4}
    \right),
\end{equation}
where $(+)$ corresponds to the results $\ket{1_0 1_5}$, and $(-)$ to $\ket{0_0 0_5}$. Passing it through the beam splitters results in
\begin{equation}
\begin{aligned}
    \frac{1}{2\sqrt{2}}\Big[ 
    & \left( 1-e^{i(\Phi\pm\delta)} \right)\left( \ket{0_1 1_2 1_3 0_4} - \ket{1_1 0_2 0_3 1_4}  \right) \\
   + & \left( 1+e^{i(\Phi\pm\delta)} \right)\left( \ket{1_1 0_2 1_3 0_4} + \ket{0_1 1_2 0_3 1_4}  \right) \Big].
\end{aligned}
\end{equation}
From the equation above we can compute the probabilities of the possible measurement results. The resulting probabilities would be conditioned on the stellar photon arrival. To recover the unconditioned probabilities, one should multiply them by $\epsilon$: the probability of the stellar photon arrival. For the ancilla qubits used for parity checks, we will associate the result $0$ with an even $(e)$ event and $1$ with an odd $(o)$ event, 
\begin{equation} \label{eq:prob11-2}
\begin{aligned}
    & p(1_1 0_2 1_3 0_4, e,e) = p(0_1 1_2 0_3 1_4, e,e)     =\frac{\epsilon}{8}\left[1 + \cos (\Phi-\delta) \right] \\
    & p(1_1 0_2 0_3 1_4, e,e) = p(0_1 1_2 1_3 0_4, e,e)  = \frac{\epsilon}{8}\left[1 - \cos (\Phi-\delta) \right] \\
    & p(1_1 0_2 1_3 0_4,o,o) = p(0_1 1_2 0_3 1_4,o,o) = \frac{\epsilon}{8}\left[1 + \cos (\Phi+\delta) \right] \\
    & p(1_1 0_2 0_3 1_4,o,o) = p(0_1 1_2 1_3 0_4,o,o)  = \frac{\epsilon}{8}\left[1 - \cos (\Phi+\delta) \right].
\end{aligned}
\end{equation}

The protocol will not work properly if the entangled ancilla photon is lost (corresponding to $0_2 0_4$ in the input state). If both stellar and entangled ancilla photons have not arrived (corresponding to the input state $\ket{0_0 0_1 0_2 0_3 0_4 0_5}$), then the following results can occur with equal probabilities:
\begin{equation} \label{eq:improper_prob}
\begin{aligned}
    &p(1_1 0_2  1_3 0_4,o,o |  \mathbf{0}_\text{star}, \mathbf{0}_\text{ancilla} )  \\
    = \ & p(0_1 1_2 0_3 1_4,o,o  |  \mathbf{0}_\text{star}, \mathbf{0}_\text{ancilla} ) \\
    = \ & p(1_1 0_2  0_3 1_4,o,o |  \mathbf{0}_\text{star}, \mathbf{0}_\text{ancilla} ) \\
    = \ & p(0_1 1_2 1_3 0_4,o,o  | \mathbf{0}_\text{star}, \mathbf{0}_\text{ancilla} ) =
    \frac{1-\epsilon}{4},
\end{aligned}    
\end{equation}
where, by $(\mathbf{0}_\text{star}, \mathbf{0}_\text{ancilla})$, we have explicitly indicated that these probabilities are conditioned on the stellar and ancilla photon not arriving. We observe that the set of possible measurement results for the properly performed procedure overlaps the set of possible results corresponding to the error of the entangled ancilla photon not arriving. We conclude that it is impossible to identify this error based on a single detection event. 

However, it should be possible to identify it if the error occurs frequently within the measurement series. Suppose that the procedure is performed correctly with probability $\eta$ and the error of not supplying the entangled ancilla photon happens with probability $(1-\eta)$. 
We assume that $\eta$ is a known parameter, since it denotes the probability of proper distribution of a known entangled state, and it should be possible to estimate it by other techniques.

Consider the results of the parity measurements. Based on (\ref{eq:prob11-2}) and (\ref{eq:improper_prob}), the probabilities of ($o,o$) and ($e,e$) events are  
\begin{equation}
\begin{aligned}
    p(o,o) & =  \frac{\eta \epsilon}{2} \\
    p(e,e) & =  \frac{\eta \epsilon}{2} + (1-\eta)(1-\epsilon).
\end{aligned}
\end{equation}
The $\eta \epsilon / 2$ terms result from the proper operation of the procedure. The $(1-\eta)(1-\epsilon)$ term results from introducing the error. We conclude that one can identify the error of the loss of the entangled ancilla photon by comparing the frequency of $(0_0 0_5)$ and $(1_0 1_5)$: if the error is introduced, then the frequency of the latter events is higher. 
After introducing the error, Eq. (\ref{eq:improper_prob}) becomes
\begin{equation} \label{eq:with_error}
\begin{aligned}
    & p(1_1 0_2 1_3 0_4, e,e) = p(0_1 1_2 0_3 1_4, e,e)     =\frac{\eta \epsilon}{8}\left[1 + \cos (\Phi-\delta) \right], \\
    & p(1_1 0_2 0_3 1_4, e,e) = p(0_1 1_2 1_3 0_4, e,e)  = \frac{\eta \epsilon}{8}\left[1 - \cos (\Phi-\delta) \right], \\
    & p(1_1 0_2 1_3 0_4,o,o) = p(0_1 1_2 0_3 1_4,o,o) \\ 
    & = \frac{\eta \epsilon}{8}\left[1 + \cos (\Phi+\delta) \right] + (1-\eta)(1-\epsilon)/4, \\
    & p(1_1 0_2 0_3 1_4,o,o) = p(0_1 1_2 1_3 0_4,o,o)  \\
    & = \frac{\eta \epsilon}{8}\left[1 - \cos (\Phi+\delta) \right]+ (1-\eta)(1-\epsilon)/4,
\end{aligned}    
\end{equation}
resulting in Fisher information
\begin{equation}
    f(\Phi)=\frac{\eta\epsilon}{2}\left( 1 + 2a \frac{\eta \epsilon - b}{a^2 - b^2} \right),
\end{equation}
where we introduced the parameters $a = 2 -2\epsilon - 2\eta +3\epsilon\eta$ and $b = \eta \epsilon \cos^2(\Phi - \delta)$. For $\eta=1$ (no error) this expression returns $f(\Phi)=\epsilon$, as it should. We note that despite introducing the error, the amount of information obtained per measurement can be still quantified by the Fisher information. 

As discussed in the main text, in the analysis of an extended source one replaces in Eq. (\ref{eq:with_error}) $\cos(\Phi\pm\delta)\rightarrow \text{Re}\left\{ \mathcal{V} e^{\mp i \delta} \right\}$, where the visibility is a complex number $\mathcal{V}=ge^{-i\Phi}$. In such case, the visibility estimation comes down to estimating the amplitude and phase of the oscillations in Eq. (\ref{eq:with_error}). 

Another way to avoid the influence of the ancilla photon loss error is performing a procedure verifying the arrival of the stellar photon before performing the NOT-based protocol. That would allow one to identify and discard the $(\mathbf{0}_\text{star}, \mathbf{0}_\text{ancilla})$ events, making the NOT-based protocol more robust against the error introduced by the loss of the ancilla photon. 

An example of a procedure verifying the stellar photon arrival is the \textit{modified quantum memory protocol} performed for one time bin, which one can consider as a subroutine performed before the NOT-based protocol. Then the NOT-based protocol enables the visibility measurement. The disadvantage of this method is that it requires distributing an additional Bell pair to the telescopes. We assume that in the future, one should be capable of doing so reliably by using a network of quantum repeaters. Moreover, that method of verification of the stellar photon arrival requires performing one more $CZ$ gate. Suppose that this gate is performed correctly with probability $\eta$ and leaves the state unchanged with probability $1-\eta$. After verifying the stellar photon arrival, the protocol can be continued only if the stellar photon is detected. That requires the presence of the stellar photon and proper operation of the $CZ$ gate. Therefore, introducing the $CZ$ gate will not introduce events that can be confused for the ones that contain valid information about the visibility. Given that the arrival of a stellar photon was detected, one continues with the original CNOT-based protocol to determine the visibility. Including the \textit{modified quantum memory protocol} before the NOT-based protocol reduces the Fisher information by a factor of $\eta$.

Finally, we should consider the events for which the stellar photon arrives and the entangled ancilla photon is lost. In this case, the results of the measurement on the $0$ and $5$ modes will be different from each other (one gets either $0_0 1_5$ or $1_0 0_5$), and such events are not taken into account when estimating the visibility. The protocol still works, but it only extracts information about the visibility from the stellar photons that arrive when an entangled ancilla is also present. The Fisher information is therefore reduced by a factor of $\eta$, which is then the probability that the arriving stellar photon will be used for parameter estimation.

\section{Multi parameter Fisher information in NOT-based protocol and Gottesman \textit{et al.} Protocol} \label{app:fisher}
We consider the elements of the Fisher information matrix for the NOT-based protocol and the Gottesman \textit{et al.} protocol. 
There are eight types of events that provide information about the visibility in the NOT-based protocol. In the case of estimating both amplitude and phase of the visibility, the event probabilities are
\begin{equation}
\begin{aligned} \label{eq:P18}
        p_1 = p_2 &  = \frac{\epsilon}{8}\left( 1 + g\cos(\Phi+\delta) \right) \\
        p_3 = p_4 &  = \frac{\epsilon}{8}\left( 1 - g\cos(\Phi+\delta) \right) \\
        p_5 = p_6 &  = \frac{\epsilon}{8}\left( 1 + g\cos(\Phi-\delta) \right) \\
        p_7 = p_8 &  = \frac{\epsilon}{8}\left( 1 - g\cos(\Phi-\delta) \right).
\end{aligned}
\end{equation}
The $g\cos(\Phi\pm\delta)$ elements were from $\text{Re}\left\{ \mathcal{V} e^{\mp i \delta} \right\} = g\cos(\Phi\pm\delta)$. Probabilities 1--4 arise from the events, where both parity checks return the \textit{odd} result. Observe that these probabilities are exactly the same as the probabilities of events that provide information about the visibility in the Gottesman \textit{et al.} protocol. Therefore, the Fisher information of their procedure is
\begin{equation} \label{eq:fisherG} 
    f_G (\Phi,g) = \sum_{k=1}^4 \frac{1}{p_k}
    \begin{pmatrix}
        \left( \frac{\partial p_k}{\partial \Phi} \right)^2 & \frac{\partial p_k}{\partial \Phi}\frac{\partial p_k}{\partial g} \\
        \frac{\partial p_k}{\partial g} \frac{\partial p_k}{\partial \Phi} & \left( \frac{\partial p_k}{\partial g} \right)^2
    \end{pmatrix},
\end{equation}
and its elements are
\begin{equation}
\begin{aligned}
    f_{G,11} & = \frac{\epsilon g^2 \sin ^2 (\Phi+\delta)}{2(1-g^2\cos^2(\Phi+\delta))}, \\
    f_{G,22} & = \frac{\epsilon \cos ^2 (\Phi+\delta)}{2(1-g^2\cos^2(\Phi+\delta))}, \\
    f_{G,12} = f_{G,21} & = \frac{\epsilon g \sin (\Phi+\delta)\cos (\Phi+\delta)}{2(1-g^2\cos^2(\Phi+\delta))}. \\
\end{aligned}
\end{equation}
The NOT-based protocol extracts information from the events where the stellar and ancilla photon landed at the same telescope, which is not the case for the Gottesman \textit{et al.} protocol. Such events return the \textit{even} results for both parity checks, and corresponds to the 5--8 indices in Eq. (\ref{eq:P18}). In other words, the NOT-based protocol extracts information from all events included in Eq. (\ref{eq:P18}), while the Gottesman \textit{et al.} protocol extracts information only from events denoted by indices 1--4. 
Calculating the Fisher information for the NOT-based protocol is done as in Eq. (\ref{eq:fisherG}), but the summation goes from 1 to 8. It results in
\begin{equation} \label{eq:FInot}
\begin{aligned}
    f_{NOT,11} & = f_{G,11} + \frac{\epsilon g^2 \sin ^2 (\Phi-\delta)}{2(1-g^2\cos^2(\Phi-\delta))}, \\
    f_{NOT,22} & = f_{G,22} + \frac{\epsilon  \cos ^2 (\Phi-\delta)}{2(1-g^2\cos^2(\Phi-\delta))}, \\
    f_{NOT,12}  & = f_{G,12} + \frac{\epsilon g \sin (\Phi-\delta)\cos (\Phi-\delta)}{2(1-g^2\cos^2(\Phi-\delta))}, \\
    f_{NOT,21}  & = f_{NOT,12},
\end{aligned}
\end{equation}
which quantifies the improvement in Fisher information over the Gottesman \textit{et al.} protocol. 

In the case of single-parameter estimation of $g$, the protocol cannot be made optimal in terms of achieving the quantum Fisher information even if one has a perfect knowledge of $\Phi$. In such case, QFI is $1/(1-g^2)$ and FI is given by the $f_{NOT,22}$ element in Eq. (\ref{eq:FInot}). QFI is saturated only when $\Phi=\delta=0$, which is almost never the case.

\section{Unmodified Quantum Memory Protocol: Example Calculation} \label{app:memory_example}

\begin{figure}
    \centering
    \includegraphics[width=8.6cm]{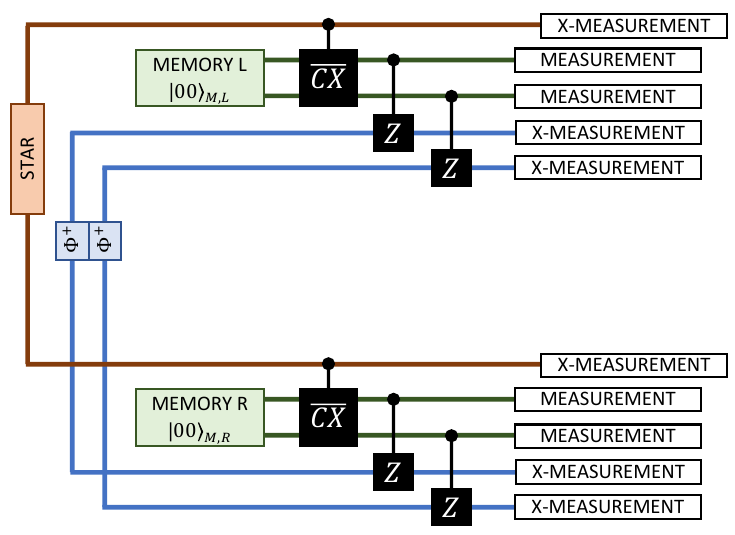}
    \caption{Schematic representation of Khabiboulline's protocol for $N=3$ time bins.}
    \label{fig:kh_protocol_scheme}
\end{figure}

We first summarize the protocol proposed by Khabiboulline \textit{et al.} The ancilla state is
\begin{equation}
\begin{aligned}
    \ket{\Psi_\text{a}} = & \ket{0...000}_\text{M,L}\ket{0...000}_\text{M,R} \otimes \\ 
    & \otimes \ket{\Phi^+}...\ket{\Phi^+}\ket{\Phi^+}\ket{\Phi^+}.    
\end{aligned}
\end{equation}
Given that the total measurement time is $T=N\tau$, we have $4\log_2(N+1)$ qubits, where $N$ is the number of time bins to be examined and $\tau$ is the width of the time bin. A quarter of them are prepared in the state $\ket{0...000}_\text{M,L}$ and located in one of the local laboratories denoted by $L$, and another quarter $\ket{0...000}_\text{M,R}$ is located in laboratory $R$. We will call them \textit{memory qubits}. The Bell pairs $\ket{\Phi^+}$ [consisting of $2\log_2(N+1)$ qubits] are distributed to the laboratories, with each laboratory receiving one qubit from each pair. The procedure for $N=3$ is summarized in Fig.~\ref{fig:kh_protocol_scheme}.

We then use a modified controlled NOT $\overline{\text{CX}}$ gate, whose action is dependent on the time bin during which the star photon arrived. The modes supplied by the star act as control qubits, and the memory provides the target qubits. The $\overline{\text{CX}}$ gate follows the pattern
\begin{equation}
\begin{aligned}
    \text{No photon arrival}: & \ket{0}\ket{0...000}_\text{M} \rightarrow                               \ket{0}\ket{0...000}_\text{M} \\
    \text{Arrival in time bin 1}: & \ket{1}\ket{0...000}_\text{M} \rightarrow                                      \ket{1}\ket{0...001}_\text{M} \\
    \text{Arrival in time bin 2}: & \ket{1}\ket{0...000}_\text{M} \rightarrow                                      \ket{1}\ket{0...010}_\text{M} \\
    \text{Arrival in time bin 3}: & \ket{1}\ket{0...000}_\text{M} \rightarrow                                      \ket{1}\ket{0...011}_\text{M} \\
        & ... \\
    \text{Arrival in time bin N}: & \ket{1}\ket{0...000}_\text{M} \rightarrow                                      \ket{1}\ket{1...111}_\text{M}.
\end{aligned}
\end{equation}
This gate performs the \textit{encoding} step: the arrival time bin is encoded in binary in the memory qubits. 

For simplicity, suppose that the star emits a photon in the third time bin and is a point source, so $\mathcal{V}=e^{-i\Phi}$, and the phase $\Phi$ is the parameter to be estimated. The state of the emitted photon is given by
\begin{equation} \label{eq:Psi_star}
    \ket{\Psi_\text{star}}=\frac{1}{\sqrt{2}}( e^{i\Phi} \ket{1_L 0_R} + \ket{0_L 1_R}).
\end{equation}

The combined state of the stellar photon and the ancilla is $\ket{\Psi_\text{star}}\otimes\ket{\Psi_\text{a}}$. Performing the $\overline{\text{CX}}$ gate results in the state
\begin{equation}
\begin{aligned}
   \ket{\Psi'} = \frac{e^{i\Phi}}{\sqrt{2}}&   \ket{1_L 0_R} \ket{0...011}_\text{M,L}\ket{0...000}_\text{M,R}  \\
    & \otimes \ket{\Phi^+}...\ket{\Phi^+}\ket{\Phi^+}\ket{\Phi^+}  \\
   + \frac{1}{\sqrt{2}} & \ket{0_L 1_R} \ket{0...000}_\text{M,L}\ket{0...011}_\text{M,R} \\ 
    & \otimes \ket{\Phi^+}...\ket{\Phi^+}\ket{\Phi^+}\ket{\Phi^+} .
\end{aligned}
\end{equation}

The next step is to perform a set of standard $\text{CZ}$ gates. Each memory qubit acts as a control and is assigned a corresponding locally available Bell-state qubit as the target (see Fig. \ref{fig:kh_protocol_scheme}). Performing the $\text{CZ}$ gates results in the state

\begin{equation}
\begin{aligned}
   \ket{\Psi''} = \frac{e^{i\Phi}}{\sqrt{2}}&   \ket{1_L 0_R} \ket{0...011}_\text{M,L}\ket{0...000}_\text{M,R}  \\
    & \otimes \ket{\Phi^+}...\ket{\Phi^+}\ket{\Phi^-}\ket{\Phi^-}  \\
   + \frac{1}{\sqrt{2}} & \ket{0_L 1_R} \ket{0...000}_\text{M,L}\ket{0...011}_\text{M,R} \\ 
    & \otimes \ket{\Phi^+}...\ket{\Phi^+}\ket{\Phi^-}\ket{\Phi^-} ,
\end{aligned}
\end{equation}
which transfers the time bin information from the memory qubits to the Bell pairs. Note that the Bell pairs form a separable state with the other qubits. One can distinguish between $\ket{\Phi^+}$ and $\ket{\Phi^-}$ by using local measurements and classical communication, since they can be rewritten in the \textit{X} basis as
\begin{equation}
\begin{aligned}
    \ket{\Phi^+} & = (\ket{+-}+\ket{-+})/\sqrt{2} \\
    \ket{\Phi^-} & = (\ket{++}+\ket{--})/\sqrt{2}.
\end{aligned}
\end{equation}
If the result of an $X$ measurement gives the same result in both laboratories, then we have the state $\ket{\Phi^-}$, otherwise we have $\ket{\Phi^+}$. This allows the parties to determine the time bin during which the photon arrived. It also allows us to determine which memory qubits were affected by the $\overline{\text{CX}}$ gate. The other memory qubits can be traced out. After measuring the Bell pairs and tracing out the irrelevant memory qubits, the analyzed state is

\begin{equation}
\begin{aligned}
   \ket{\Psi'''} = \frac{1}{\sqrt{2}} \Big( 
    & e^{i\Phi} \ket{1_L 0_R}\ket{11}_\text{M,L} \ket{00}_\text{M,R} \\
    & + \ket{0_L 1_R} \ket{00}_\text{M,L} \ket{11}_\text{M,R} \Big).
\end{aligned}
\end{equation}
The star photon mode is decoupled from the memories by the measurement in the \textit{X} basis. In any order, all but one of the memory qubits are measured in the \textit{X} basis, and the final memory qubit is measured in the rotated basis spanned by $\ket{\pm_\delta} = \frac{1}{\sqrt{2}}(\ket{0}\pm e^{i\delta}\ket{1})$. If $n_-$ denotes the number of times the $X$ measurements return the  $\ket{-}$ result, then the probabilities of the measurement results in the rotated basis are
\begin{equation} \label{eq:p1}
    P(\pm_\delta) = \frac{1}{2}\left[ 1 \pm (-1)^{n_-}\cos(\Phi+\delta) \right];
\end{equation}
for an extended source, this becomes
\begin{equation} \label{eq:p2}
     P(\pm_\delta) = \frac{1}{2}\left[ 1 \pm (-1)^{n_-}
     \text{Re}\left\{ \mathcal{V} e^{-i\delta} \right\} \right].
\end{equation}

\section{Modified Quantum Memory Protocol: Example Calculation} \label{app:memory_example_mod}

Suppose the star provides a photon in the state (\ref{eq:Psi_star}), which arrives at the telescopes in the third time bin. The combined state of the star photon and the ancilla qubits is
\begin{equation}
    \frac{1}{\sqrt{2}}(e^{i\Phi}\ket{1_L 0_R} + \ket{0_L 1_R}) \otimes
    \ket{\Phi^+}...\ket{\Phi^+}\ket{\Phi^+}\ket{\Phi^+}.
\end{equation}
Performing the modified controlled phase gate results in
\begin{equation}
    \frac{1}{\sqrt{2}}(e^{i\Phi}\ket{1_L 0_R} + \ket{0_L 1_R}) \otimes
    \ket{\Phi^+}...\ket{\Phi^+}\ket{\Phi^-}\ket{\Phi^-}.
\end{equation}
Both laboratories measure the ancilla qubits in the \textit{X} basis and establish the time bin during which the star photon arrived. After these measurements, the star photon is left in the state (\ref{eq:Psi_star}). The stellar photon provides us with two single-rail qubits which we can rewrite in different bases.
Rewriting the state of the qubit in laboratory $L$ in the \textit{X} basis and the qubit in $R$ in the rotated basis gives
\begin{equation}
\begin{aligned}
    \ket{\Psi_\text{star}}=\frac{1}{\sqrt{2}} 
     \Big[ & \cos\left(\frac{\Phi+\delta}{2}\right) (\ket{+,+_\delta}-\ket{-,-_\delta}) \\
     + i & \sin\left(\frac{\Phi+\delta}{2}\right) (\ket{+,-_\delta}-\ket{-,+_\delta}) \Big],
\end{aligned}
\end{equation}
which results in the probabilities \begin{equation}
\begin{aligned}
    P(+,+_\delta) = P(-,-_\delta) = & \frac{1}{4} \left[1 + \cos(\Phi + \delta) \right] \\
    P(+,-_\delta) = P(-,+_\delta) = & \frac{1}{4} \left[1 - \cos(\Phi + \delta) \right].
\end{aligned}
\end{equation}

\section{Modified Quantum Memory Protocol: Example of Resource Estimation} \label{app:resources}

Suppose that one needs to examine the state of stellar modes arriving within $N=7$ time bins. In that case, the protocol requires three Bell states to perform the binary encoding. If the star photon arrives within the third time bin, then the ancilla is modified to
\begin{equation} 
    \ket{\Phi^+}\ket{\Phi^+}\ket{\Phi^+} \rightarrow
    \ket{\Phi^+}\ket{\Phi^-}\ket{\Phi^-},
\end{equation}
where the first two Bell states have been flipped in accordance with the binary representation of $n=3$ ($011$). The states $\ket{\Phi^+}$ and $\ket{\Phi^-}$ can be distinguished by local measurements and classical communication by measuring both qubits in the $X$ basis; if the results are the same, then the measured state is $\ket{\Phi^-}$, otherwise it is $\ket{\Phi^+}$. Applying the same procedure to all ancilla pairs returns the time bin during which the star photon arrived. If no photon arrived, then the ancilla remains unchanged. The next step is to measure the star photon modes for all time bins in the $X$ (mode 1) and rotated (mode 3) bases. After both laboratories complete the measurement on the ancilla, they communicate the result to each other to establish within which time bin the stellar photon has arrived. For that time bin, the possible stellar photon mode measurement results are described by the probabilities (\ref{eq:p1}) and (\ref{eq:p2}). These probabilities saturate the QFI in the case of a measurement of a point source, i.e., the phase measurement. The same is not true for an extended source, for the reasons discussed before.

For example, if $N=100$, then one needs $100$ distributed entangled pairs to perform a unary protocol, but only $7$ entangled pairs to perform the \textit{NOT-based protocol}, the same scaling as in the protocol of Khabiboulline \textit{et al.} Of course, in a practical stellar interferometry experiment, one needs to determine the coherence between the stellar photon modes at many baselines, and each coherence measurement requires analysis of 100--1000 photons. If we consider stellar photons with bandwidth of 10 GHz, there will be $\sim$10\textsuperscript{6} of these (around 600 nm) every second ($\sim$10\textsuperscript{8} temporal modes with an occupancy of $\sim$0.01, comparable to a typical blackbody source in visible light range), or $\sim$10\textsuperscript{4} in the $\sim$10 ms data acquisition time needed to beat turbulence fluctuations of the relative phase. Thus, our terrestrial source will need to produce and distribute matching-bandwidth photons at a rate $\sim$10\textsuperscript{8} s\textsuperscript{-1} in order to have 100 coincidental events in 10 ms. This is definitely challenging, but not infeasible.

\bibliography{apssamp}

\end{document}